\documentclass[apl,twocolumn,showpacs,amsmath,amssymb,superscriptaddress]{revtex4-1}
\usepackage{epsf}
\usepackage{graphicx}
\usepackage{bm}
\usepackage{amsmath}
\usepackage{color}
\usepackage{notes2bib}
\usepackage{epstopdf}

\bibnotesetup{
note-name = , use-sort-key = false }
\begin{document}
\title{Resonant indirect exchange via remote 2D channel}

\author{I.~V.~Rozhansky}
\email{rozhansky@gmail.com} \affiliation{Ioffe Institute, 194021
St.Petersburg, Russia} \affiliation{Lappeenranta University of
Technology, FI-53851 Lappeenranta, Finland}
\affiliation{ St. Petersburg State Polytechnic University, 195251 St. Petersburg, Russia}
\author{I.~V.~Krainov}
\affiliation{Ioffe Institute, 194021 St.Petersburg, Russia}
\author{N.~S.~Averkiev}
\affiliation{Ioffe Institute, 194021 St.Petersburg, Russia}
\author{B.~A.~Aronzon}
\affiliation{P.N. Lebedev Physical Institute, 119991 Moscow, Russia}
\affiliation{
National Research Centre "Kurchatov Institute", 123182 Moscow, Russia}
\author{A.~B.~Davydov}
\affiliation{P.N. Lebedev Physical Institute, 119991 Moscow, Russia}
\author{K.~I.~Kugel}
\affiliation{Institute of Theoretical and Applied Electrodynamics,
125412 Moscow, Russia}
\author{V.~Tripathi}
\affiliation{Tata Institute of Fundamental Research, Mumbai 400005, India}
\author{E.~L\"ahderanta}
\affiliation{Lappeenranta University of Technology, FI-53851 Lappeenranta, Finland}

\begin{abstract}
We apply the resonant indirect exchange 
interaction theory to explain the ferromagnetic properties of the hybrid heterostructure
consisting of a InGaAs-based quantum well (QW) sandwiched between GaAs barriers with a remote Mn delta-layer. The experimentally obtained dependence of the Curie temperature on the QW depth exhibits a maximum related to the region of resonant indirect exchange. We suggest the theoretical explanantion and a fit to this dependence as a result of the two contributions to ferromagnetism - the intralayer contribution and the resonant exchange contribution provided by the QW.

\end{abstract}


\date{\today}

\maketitle

\section{Introduction}
Dilute magnetic semiconductors (DMS) have been attracting a lot of attention
for quite a while \cite{Jungwirth2014}.
A lot of efforts have been put forward
to combine the numerous advantages of semiconductors with the spin-related phenomena
introduced by the magnetic impurities. In this field, however, still much remains unclear.
For instance, the details of the mechanism responsible for the ferromagnetic properties of
GaAs doped with a small amount of Mn has not yet been clarified \cite{Jungwirth}. It is commonly accepted
that the ferromagnetism in (Ga,Mn)As is due to the indirect exchange interaction mediated by the holes.
The highest Curie temperature achieved for bulk dilute (Ga,Mn)As samples is near $200K$ which is still far below the room temperature \cite{Dietl2014}. While it is the Mn solubility limit that basically prevents the increase of $T_c$ in bulk samples \cite{Ohno1999}, the indirect exchange depends on the concentration of the holes. In this regard the GaAs
heterostructures with a Mn layer coupled to a remote 2D holes channel have gained a considerable interest
\cite{Rupprecht,Nishitani,Dorokhin2014,Aronzon2010}.
It has been demonstrated that GaAs heterostructure with a Mn $\delta$-layer located in a vicinity of In$_x$Ga$_{1-x}$As quantum well (QW) shows ferromagnetic behavior similar to that of the bulk Mn-doped GaAs DMS. It was discovered, however, that the dependence of the Curie
temperature on the QW depth shows a non-monotonic behavior
\cite{Aronzon2013}. It was suggested that the non-monotonic
behavior originates from falling of the hole bound state at Mn ion
 into the energy range of occupied 2D heavy holes subband of the first QW size
quantization level \cite{OURPSS2014}. A theory of the indirect exchange via a remote conducting channel was
developed in \cite{OURPSS2014},\cite{OURPRB2013}. The theory predicted enhancement of the exchange interaction strength due
to resonant tunnel coupling of a bound state at magnetic ion with the continuum of delocalized 2D states in the channel.
In this paper we present a
comparison with the experimentally observed dependence
of the Curie temperature on the QW depth and analyze how the temperature affects the indirect exchange interaction in the resonant case.
As soon as we are talking about 2D structures, the QW and the Mn
$\delta$ - layer, one should be aware of what is meant by the Curie temperature as
there is no possibility of spontaneous breaking of a continuous (rotational) symmetry in our 2D Heisenberg ferromagnet.
In our theoretical considerations we actually consider the mean effective exchange constant, i.e. the energy of the indirect exchange interaction between the two neighbouring Mn ions. For a 3D bulk case it is indeed close to the critical Curie temperature for the system to undergo the ferromagnetic phase transition. In the 2D case, however to leave the things consistent it is reasonable to define the Curie temperature as the one marking the onset of 'local ferromagnetic order', when the magnetic correlation length well exceeds the distance between the Mn ions.
In the experiment the so defined Curie temperature is obtained from a maximum (bump) on the dependence of
in-plane electrical resistance vs temperature which is known to be related to the onset of the ferromagnetic order \cite{Res1},\cite{Res2}.
For more detailed discussion on the critical temperature in the system under study see Ref.\cite{Aronzon2011},\cite{Aronzon2012}.
\section{Theory of resonant indirect exchange}
We consider the problem of resonant indirect exchange between two magnetic ions $i$ and $j$ via a remote 2D channel as shown schematically in Fig 1a.
\begin{figure}[t]
  \leavevmode
 \centering
\includegraphics[width=0.5\textwidth]{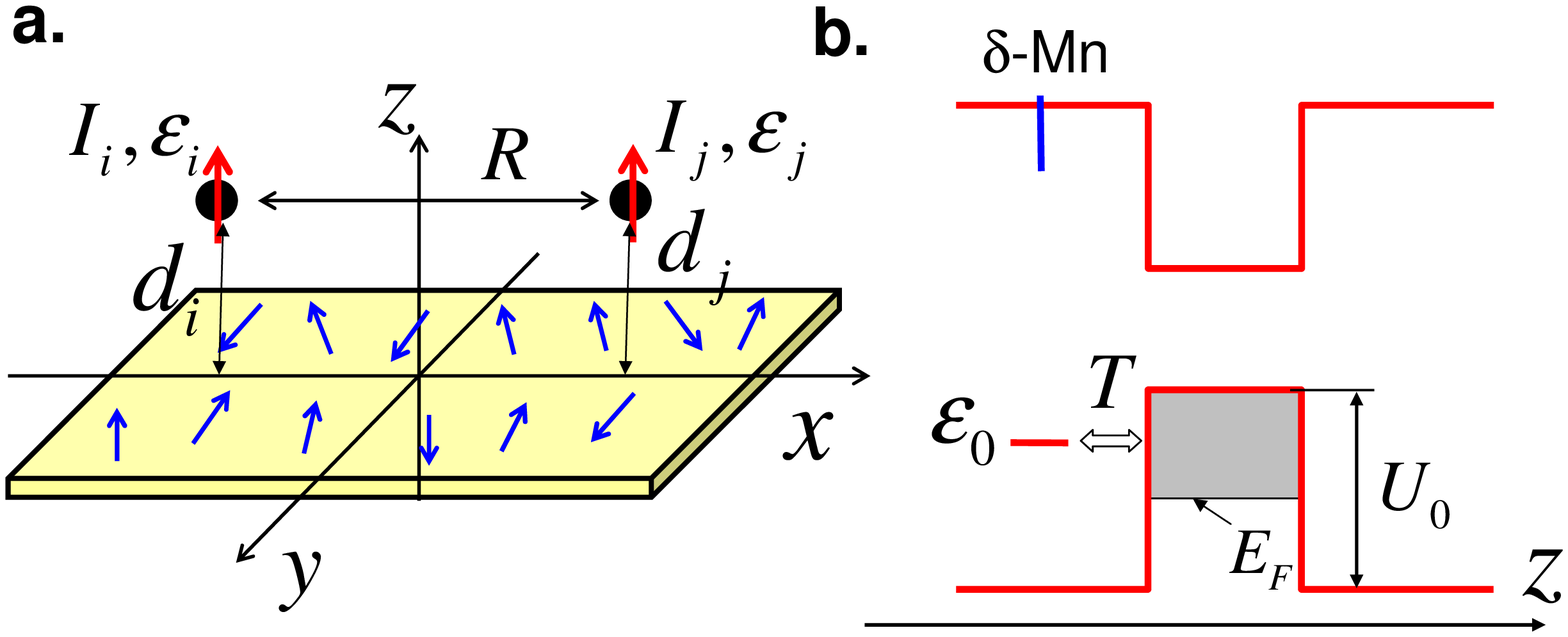}
 \caption{The illustration of the indirect exchange interaction via remote channel (a) and band diagram of a InGaAs-based heterostructure with a QW and a remote Mn $\delta$-layer (b).}
 \label{figScheme}
\end{figure}
Here $I_{i,j}$ is the spin projection of the $i$th($j$th) magnetic ion, $\varepsilon_{i,j}$ is the ions' bound state energy, $d_{i,j}$  the distances between the ions and the channel.
Indirect exchange interaction mediated by free carriers is usually described on the basis of Ruderman-Kittel-Kasuya-Yosida (RKKY) theory \cite{Kittel}.  Separation of the magnetic ions from the free carriers gas by a potential barrier leads to the suppression of the effect due to weak penetration of the 2D carriers wavefunction into the region containing magnetic centers. However, if a magnetic ion possess a bound state having the energy within the 2D gas energy spectra a resonant tunneling may occur. The resonant coupling of a bound state with the 2D continuum prevents the problem to be straightforwardly attacked with RKKY approach.
The Hamiltonian of the two magnetic ions $i,j$ coupled to the free electron gas can be expressed in the tunneling Hamiltonian formalism:
\begin{equation}
\label{eqH} H = H_0  + H_T  + H_J, \end{equation} where $H_0$ -- the
Hamiltonian of the system without tunnel coupling and spin-spin
interaction, $H_T$ -- the tunnelling term,
$H_J$ -- the exchange interaction term. In the second quantization
representation:
\begin{align}
 H_0  &=
\varepsilon _i a_i ^ +  a_i  + \varepsilon _j a_j ^ +  a_j  + \int
{\varepsilon _\lambda  c_{\lambda}^ +  c_{\lambda } } d\lambda,
  \nonumber\\
 H_T  &=
\int {\left( {t_{i\lambda }a_i^+  c_{\lambda }   + t_{j\lambda }
a_j^+ c_{\lambda }    + h.c.} \right)d\lambda },
  \nonumber \\
 H_J  &= JA\left( {I_i sa_i^ +  a_i  + I_j sa_j^ +  a_j } \right) ,
  \end{align}
where $a^+_i,a_i$ are the creation and annihilation operators
for the bound states at the impurity ion $i$, characterized by
the energy level $\varepsilon_i$ and localized wavefunction
$\psi_i$,  $c^+_{\lambda}, c_{\lambda}$ are the creation and
annihilation operators for a continuum state characterized by the
quantum number(s) $\lambda$, having the energy $\varepsilon
_{\lambda}$ and the wavefunction $\varphi_{\lambda}$, $J$ is the exchange constant, A is the squared wavefunction amplitude
at the ions site, $s$ is the 2D carrier's spin projection,
 $t_{i,\lambda}$ is the Bardeen's tunneling matrix element given by \cite{OURPSS2014}:
\begin{equation}
\label{eqtunpar} t_i \left( k \right) = \sqrt{\frac{{\hbar ^2 T_i
}}{{2\pi m}}}e^{  i{\bf{kR}}_{{\bf{i}}} } ,
\end{equation}
where m is the 2D continuum density of states effective mass, $T_i$ is the energy parameter for the tunneling:
\begin{equation}
T_i =  \alpha U_0 e^{ - 2qd_i},
\end{equation}
where $U_0$ is the height of the potential barrier separating the magnetic centers from the channel,
$q=\sqrt{2m_{\perp}U_0}/\hbar$,
$m_{\perp}$ being an effective mass in the direction of the tunneling,
the dimensionless parameter $\alpha$ depends on the channel and magnetic centers details \cite{OURPRB2013}.
We obtained the exchange interaction energy between the two ions in the form \cite{OURPSS2014}, \cite{OURJMMM2014}:
\begin{equation}
\label{eqExchEnergy}
E_{ij} = \frac{1}{\pi} \int_0^{E_F} \,d\varepsilon \ \arctan\left[ \frac{8\pi^2 j^2 T_i T_j J_0 (k R)Y_0(kR)}{\left((\varepsilon_i - \varepsilon)^2-j^2\right)\left((\varepsilon_j - \varepsilon)^2-j^2\right)} \right],
\end{equation}
where $j=J A |I| |s|$ is the exchange interaction strength,
$E_F$ denotes the Fermi level of the carriers in the channel (zero temperature is assumed),
$k=\sqrt{2m\varepsilon}/\hbar$,
$J_0, Y_0$ are Bessel and Neumann functions of zeroth order,

The formula (\ref{eqExchEnergy}) accounts both for the case of resonant and non-resonant tunnel coupling between the magnetic ions and the channel.
The resonant case corresponds to the bound states energy lying within the energy range of the occupied states in the 2D channel:
\begin{equation}
\label{eqResCond}
{\varepsilon _i},{\varepsilon _j} \in \left[ {0,{E_F}} \right].
\end{equation}
In this regime the main contribution to the exchange energy (\ref{eqExchEnergy}) comes from the poles of the arctangent argument
and can be estimated as
\begin{align}
\label{eqTi}
& E_{ij}\approx\gamma\sqrt {j T},&\text{if }\beta>1 \nonumber \\
& E_{ij}\approx\gamma \beta \sqrt {j T},&\text{if }\beta<1 \nonumber \\
& \beta  = \frac{{\sqrt {j T} }}{{\left| {{\varepsilon _i} - {\varepsilon _j}} \right|}},\,\,\, T = \sqrt {{T_i}{T_j}}.
\end{align}
Here $\gamma$ is given by:
\begin{equation}
\label{eqgamma}
\gamma  = \sqrt {2\pi } \left[ {J_0 \left( {k_i R_{ij} } \right)Y_0 \left( {k_i R_{ij} } \right) + J_0 \left( {k_j R_{ij} } \right)Y_0 \left( {k_j R_{ij} } \right)} \right]^{1/4},
\end{equation}
where $R_{ij}$ is the distance between the ions.
$\gamma$ is the parameter that incorporates the oscillating behavior of the indirect exchange in the similar way
that standard RKKY theory does. Unlike RKKY theory, here the Fermi wavevector $k_F$ is
replaced by the 'resonant' wavevectors corresponding to the bound levels:
$k_i=\sqrt{2m\varepsilon_i}/\hbar$.
For the experimental situation considered below the parameter $\gamma$ appears to be $\gamma\approx 1$ being still far from the first maximum of the oscillations.
The approximation (\ref{eqTi}) is quite good as illustrated in Fig. 2. The dotted curve shows the exchange interaction energy
calculated according to (\ref{eqExchEnergy}) for the case $\varepsilon_i=\varepsilon_j=\varepsilon_0$, while the solid curve shows the approximation
(\ref{eqTi}) assuming $\gamma=1$. 
The value of the exchange energy in the resonant case is much larger than in the non-resonant one.
The latter case agrees well with the RKKY approach, the integration (\ref{eqExchEnergy}) for the case when the arctangent argument has no poles and
therefore is small as far as the tunneling is weak yields \cite{OURPSS2014}:
\begin{align}
\label{eqNonRes}  &E_{nr} = \frac{{8\pi T^2 j^2 E_F }}{{\varepsilon
_0 ^4 }}\chi\left(R\right), \nonumber\\
&\chi \left( R \right) = J_0 \left( {k_F R} \right)Y_0 \left( {k_F
R} \right) + J_1 \left( {k_F R} \right)Y_1 \left( {k_F R} \right),
\end{align}
where $R$ denotes the mean distance between the ions.
Note that the resonant and non-resonant cases have different parametric dependence on the tunneling parameter $T$ and the
exchange parameter $j$, this leads to substantial amplification of the indirect exchange in the resonant case.
\begin{figure}
 \centering
 \includegraphics[width=0.4\textwidth]{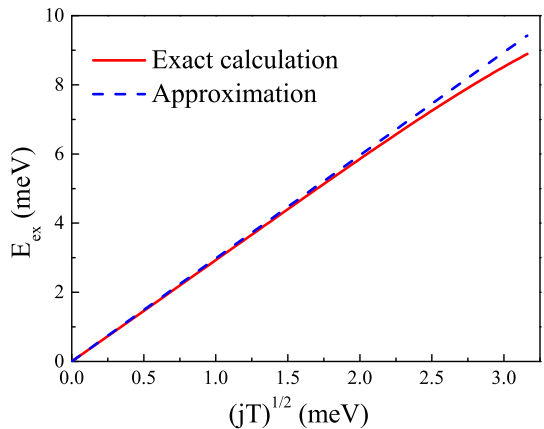}
 \caption{Approximation for indirect resonant exchange energy (dashed line) compared to
exact result (solid line).}
 \label{figTj}
\end{figure}
\section{Comparison with experiment}
We applied the theory of the resonant exchange to the hybrid
InGaAs-based semiconductor hetterostructure doped with a $\delta$-layer of Mn, which have been also studied experimentally.
The energy diagram for the system under study is shown schematically in Fig.1b.
It is a GaAs/Mn $\delta$-layer/GaAs/In$_x$Ga$_{1-x}$As/GaAs heterostructure. The Mn content (Mn layer effective thickness) is 0.25-0.3 monolayers (ML), the spacer thickness $d$, between the Mn layer and the In$_x$Ga$_{1-x}$As quantum well (QW) is $d=3$ nm,
the thickness of the QW is  10 nm and its depth is controlled by In concentration $x$. For detailed description of the structure see \cite{Aronzon2010}.
The samples were shown to exhibit ferromagnetic properties, which were found to be non-monotonously  dependent on the QW depth \cite{Aronzon2013}. The
Curie temperature derived from the resistance anomaly appears to depend
on the parameters of the QW thus favouring the idea
that the indirect exchange interaction is partly due to 2D holes sitting in the QW.
The interaction between the Mn ions mediated by the 2D holes in the QW must be considered with account for the resonant indirect exchange, because
the acceptor binding energy of Mn in GaAs is comparable to the QW depth for the holes. This makes it possible to meet the resonant condition (\ref{eqResCond}). This very case is shown in Fig. 1b,
where $\varepsilon_0$ denotes the average energy of the bound state of a hole at Mn in the $\delta$-layer (zero energy corresponds to the first heavy holes quantization level in the QW).
It is also worth noting that in real samples the the Mn $\delta$-layer has a certain width. In fact due to the Mn ions diffusion its halfwidth is known to be around 1.5 nm \cite{AronzonKovalchuk}.
It is thus natural to expect that the exchange via QW is not the only contribution, i.e. without QW the ferromagnetic properties of the Mn layer would be resembling the bulk ferromagnetism of a dilute magnetic semiconductor sample of a small yet finite thickness around 3 nm.
The ferromagnetic properties of the Mn $\delta$ layer embedded into GaAs
matrix has been also studied theoretically \cite{Tugushev2009}.
\begin{figure}
 \centering\includegraphics[width=0.4\textwidth]{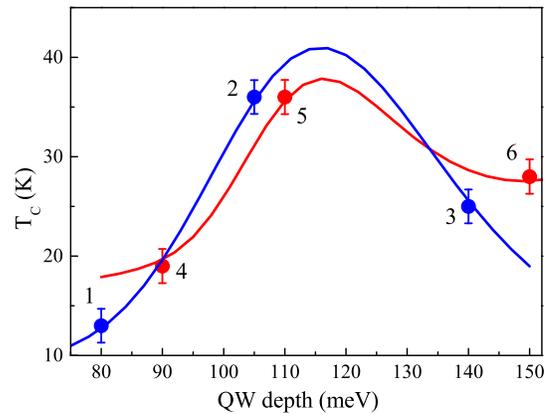}
 \caption{Dependence of the Curie temperature on the QW depth. Experimental data (circles)
 and the theoretical fit (solid curves).}
 \label{figFit}
\end{figure}
We made a fit to the experimental data \cite{Aronzon2013} (circles in Fig.3) assuming that
there are two contributions to the indirect exchange interaction between the Mn ions. The first one is
assumed to be itinerant ferromagnetism of the Mn $\delta$-layer itself due to the weakly localized holes
located primarily in this layer in the same manner as the ferromagnetism in the bulk dilute GaMnAs
semiconductor is believed to emerge. This contribution does not depend on the QW properties.
 The second contribution is the resonant indirect exchange via the 2D holes of the QW.
This one is treated on the basis of our theoretical result (\ref{eqExchEnergy}).
We demonstrate that the maximum of the Curie temperature corresponds to the resonant indirect exchange via the 2D holes of the QW while its decrease for both too shallow or too deep QW is explained by driving the system out of the resonance (\ref{eqResCond}).
In our calculation we assumed the Curie temperature being the sum of the two terms:
\begin{equation}
\label{eqTc12}
T_C  = T_{C1}  + T_{C2},
\end{equation}
where $T_{C1}$ does not depend on the QW properties.
In calculation of the second term  $T_{C2}$ we assumed that the energy levels at Mn ions are
normally distributed having an average value $\varepsilon_0$ and dispersion $\sigma_\varepsilon$.
The
distance between the neighbouring ions was assumed to be constant, equal to the mean one $R$.
For the Mn $\delta$-doping of 0.3 ML one can take $R=1.5$ nm.
We checked that taking into account some distribution over the distances as well as varying the mean value
has little effect on the resonant exchange term. This is because $\gamma$ defined by (\ref{eqgamma}) is a very weak
function of $R$ in the vicinity of $\gamma=1$.
On the contrary, the Mn bound levels energy distribution does play an important role and
must be accounted for.
In order to introduce the energy levels distribution into the fitting procedure it is convenient to
replace the approximation (\ref{eqTi}) by a similar function:
\begin{equation}
\widetilde{E}(\varepsilon_i,\varepsilon_j) = \frac{\gamma^2 j T}{|\varepsilon_i-\varepsilon_j| + \gamma \sqrt{j T}}
\end{equation}
The resonant contribution to the Curie temperature $T_{C2}$ (\ref{eqTc12}) is calculated using the following expression:
\begin{equation}
\label{eqTc2}
T_{C2} = \frac{2}{k_B} \int_0^{E_F} d\varepsilon \ \int_{-\infty}^{+\infty}d\varepsilon' \ P(\varepsilon) P(\varepsilon') \widetilde{E}(\varepsilon,\varepsilon'),
\end{equation}
where:
\begin{equation}
P(\varepsilon) = \frac{1}{\sqrt{2 \pi} \sigma_\varepsilon} e^{-\frac{(\varepsilon-\varepsilon_0)^2}{2 \sigma_\varepsilon^2}},
\end{equation}
$k_B$ is the Boltzmann constant.
In the limiting case of a delta-like bound states energy distribution $\sigma_\varepsilon\rightarrow0$
the expression (\ref{eqTc2}) yields:
 \begin{equation}
T_{C2}  = \left\{ {\begin{array}{*{20}c}
   {\frac{2}{{k_B }}\gamma \sqrt {jT} ,\,\,\varepsilon _0  \in \left[ {0,E_F } \right]}  \\
   {0,\,\,\,\text{otherwise}}
\end{array}} \right.,
\end{equation}
i.e. the resonant contribution vanishes
whenever $\varepsilon_0$ goes below or above the energy range occupied by the carriers in the QW.
The approach was used to fit the experimental values of the Curie temperature measured
for two series of samples. The samples 1-3 had 0.25 ML of Mn and the QW depth for the holes $U_0$
varied from 80 to 140 meV, the samples 4-6 had 0.3 ML of Mn with $U_0$ varied from 90 to 150 meV.
The two theoretical fits for the two series of experimental points are presented in Fig.\ref{figFit}.
The average value of the Mn bound state for the best fit was $\varepsilon_0=U_0-103$ meV, i.e. $103$ meV above the top of the valence band for GaAs, which roughly matches the Mn acceptor binding energy ($\approx110$ meV). This value was the same for the two fits.
The holes concentration and thus the holes Fermi level was derived independently from the transport experiments \cite{Aronzon2010},\cite{Zaitsev2012} and the QW depth from the optical experiments.
These values are given in the Table 1 along with the other parameters of the fit.
Along with the mean bound state energy the fitting parameters were the non-resonant component of the Curie temperature $T_{C1}$, the bound state energy dispersion $\sigma_\varepsilon$ and the product $jT$, assumed
to be the same for all the ions in the layer.
As it is seen from Table 1, for the fit covering samples 4-6 $T_{C1}$ appeared to be higher than for the samples 1-3, this is consistent with the latter having weaker Mn doping. The larger $jT$ product for
the more heavily doped samples can be understood if we recall that the so-called $\delta$ -- layer of Mn in reality has rather thick spatial distribution, expected to be thicker for larger Mn concentration due to Mn diffusion \cite{Aronzon2010}. Thus, the minimum distance between the Mn layer and the QW is expected to decrease with increase of Mn doping level resulting in increase of the tunneling parameter. The difference in the energy levels dispersion $\sigma_\varepsilon$ perhaps cannot be unambiguously explained with the similar plain arguments. What is
read from the fit (Table 1) is that the diagonal disorder is somewhat smaller in the Mn layer with higher Mn concentration. To summarize we conclude that the fit demonstrates good agreement with the experimental data and the obtained values of the fitting parameters seem quite reasonable. Thus, the experimental data can be described by the two contributions as explained above.
\begin{table}
  \centering
\begin{tabular}{|c|c|c|c|c|c|c|c|}
  \hline
   No& $T_c$, K & $U_0$, meV & p, cm$^{-2}$ & $E_F$, meV & $T_{C1}$, K & $\sqrt{jT}$, meV & $\sigma_\varepsilon$, meV  \\
  \hline
  1 & 13 & 80 & $5.6 \cdot 10^{11}$ & 7.8 & 9 & 3.3 & 18 \\
  2 & 36 & 105 & $8.9 \cdot 10^{11}$ & 12.5 &9 &3.3 & 18 \\
  3 & 25 & 140 & $1.8 \cdot 10^{12}$ & 25.2 &9 & 3.3 & 18 \\
  4 & 19 & 90 & $0.7 \cdot 10^{11}$ & 1.0 & 17 & 4.1 & 13 \\
  5 & 36 & 110 & $3.0 \cdot 10^{11}$ & 4.2 & 17 &4.1 & 13 \\
  6 & 28 & 150 & $2.3\cdot10^{12}$ & 32.2 & 17 & 4.1 & 13 \\
  \hline
\end{tabular}
  \caption{The parameters of the fit}\label{table1}
\end{table}
\begin{figure}
 \centering\includegraphics[width=0.4\textwidth]{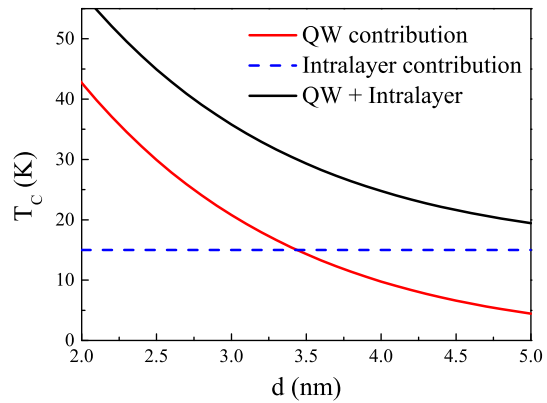}
 \caption{The calculated dependence of the Curie temperature on the distance between
 QW and Mn $\delta$--layer.}
 \label{figTcompare}
\end{figure}
    It might be also useful to illustrate the interplay between the two contributions to the ferromagnetism. Shown in Fig. \ref{figTcompare} is the dependence of the two contributions
    as a functions of the distance $d$ between the Mn $\delta$ -- layer and the QW.
 $T_{C1}$ term is referred as intralayer contribution in Fig. \ref{figTcompare}. It is, of course, independent of $d$.  The QW contribution does depend on $d$ through the tunneling parameter.
 For the tunneling case this dependence appears to be weaker than for the non-resonant one as
 $T_C$ roughly follows the $\sqrt{T}$ dependence (\ref{eqTi}) rather than $T^2$ (\ref{eqNonRes}).
 The illustration presented in Fig. \ref{figTcompare} corresponds to the parameters of the sample 5 (Table 1) and extrapolated to spacer thickness other than that of sample 5 ($d=3$ nm). We note here that
 the dependence of $T_C$ on $d$ appears to be still too strong compared to the experimental observations
  \cite{Aronzon2013}. We attribute this disagreement to the
uncertainty in determination of the distance $d$ between Mn layer and QW due to the finite thickness of the Mn layer being around 3 nm.
  The detailed analysis here requires more experimental data.
\section{Summary}
On the basis of the the previously developed theory of the resonant indirect exchange
interaction we analyzed the ferromagnetic properties of the hybrid heterostructure
consisting of a InGaAs QW and remote Mn layer. The experimentally observed non-monotonous
dependence of the Curie temperature on the QW depth was explained and fit as
the result of two contributions to ferromagnetism. The first one is the intralayer contribution
stems from the same mechanism as that of the ferromagnetism in bulk dilute GaMnAs samples.
The second contribution is the resonant indirect exchange via the 2D holes populating the QW.
It is this mechanism that is responsible for the non-monotonous behavior of $T_C$.
As only the second contribution depends on the distance between the QW and Mn layer, further
experimental investigations are required to separate the two mechanisms.
\section{Acknowledgements}
IVR, IVK, NSA acknowledge support of Russian Science Foundation (project 14-12-00255),
BAA, KIK, ABD, VT are grateful for support to Indian-Russian collaborative grant DST-MSE 14.513.21.0019, RFBR grant 14-02-00586,
VT acknowledges support of DST for Indo-Russian collaborative grants RFBR-P-141 and RMES-02/14.
\bibliography{FanoExchange}

\end{document}